%% file: povarov1.tex
\documentclass[aps,prl,reprint,superscriptaddress]{revtex4-1}
\usepackage{amsmath,amssymb,bm}
\usepackage{graphicx}
\usepackage{epstopdf}
\usepackage{latexsym}
\usepackage{subfigure}
\usepackage{color}
\usepackage{hyperref}

\newcommand{\ccc}{{Cs$_2$CuCl$_4$}}

\newcommand{\be}{\begin{equation}}
\newcommand{\ee}{\end{equation}}
\newcommand{\bea}{\begin{eqnarray}}
\newcommand{\eea}{\end{eqnarray}}

\begin{document}

\title{Modes of magnetic resonance in the spin liquid phase of Cs$_{2}$CuCl$_{4}$}

\author{K. Yu. Povarov}
    \email{povarov@kapitza.ras.ru}
    \affiliation{P. L. Kapitza Institute for Physical Problems, RAS, 119334 Moscow, Russia}
\author{A. I. Smirnov}
    \affiliation{P. L. Kapitza Institute for Physical Problems, RAS, 119334 Moscow, Russia}
    \affiliation{Moscow Institute for Physics and Technology , 141700, Dolgoprudny, Russia}
\author{O. A. Starykh}
    \affiliation{Department of Physics and Astronomy, University of Utah, Salt Lake City, Utah 84112, USA}
    \affiliation{Max-Planck-Institut f\"ur Physik Komplexer Systeme, D-01187 Dresden, Germany}
\author{S. V. Petrov}
    \affiliation{P. L. Kapitza Institute for Physical Problems, RAS, 119334 Moscow, Russia}
\author{A. Ya. Shapiro}
    \affiliation{A.V.Shubnikov Institute of Crystallography, RAS, 119333 Moscow, Russia}

\date{\today}

\begin{abstract}
We report the observation of a  frequency shift and splitting of
the electron spin resonance (ESR) mode of the low-dimensional
$S=1/2$ frustrated antiferromagnet Cs$_{2}$CuCl$_{4}$ in the
spin-correlated state below the Curie-Weiss temperature 4 K but
above the ordering temperature 0.62 K. The shift and splitting
exhibit strong anisotropy with respect to the direction of the
applied magnetic field and do not vanish in zero field.
The low-temperature evolution of spin resonance response is a result of the specific modification
of one-dimensional spinon continuum under the action of the uniform Dzyaloshinskii-Moriya interaction (DM)
within the spin chains.
Parameters of the uniform DM interaction are derived from the experiment.
\end{abstract}


\maketitle


Quantum antiferromagnets on triangular lattice exhibit a variety of unusual  phases with intriguing physical
properties. In this work we report results of electron spin resonance (ESR) study of quasi two-dimensional (2D)
triangular antiferromagnet \ccc\ in the paramagnetic phase. \ccc\ realizes distorted triangular lattice with
strong exchange $J$ along  the base of the triangular unit and weaker exchange integral $J^\prime = 0.34 J$
along lateral sides of the triangle. The material orders below $T_{\rm N}=0.62$ K into a long-range 3D spiral
state \cite{ColdeaPRL}. This N\'eel point is far below the Curie-Weiss temperature $T_{CW}$=4 K. In the
intermediate temperature range $T_{\rm N} < T< T_{\rm CW}$ a spin-liquid phase with strong in-chain spin
correlations takes place. Magnetic properties of \ccc\ are well described by a quasi-one-dimensional (1D) model
of weakly coupled spin $S=1/2$ chains.  In particular, inelastic neutron scattering experiments
\cite{ColdeaPRL,ColdeaPRB} are nicely explained by a parameter-free theory of dispersing continuum of
one-dimensional spinons \cite{kohno2007}. Moreover, the same description, extended  in Ref. \cite{Starykh} to
include residual symmetry-allowed Dzyaloshinskii-Moriya (DM) interactions between the Cu-spins, is able to
consistently explain majority of the phases in the $T-H$ phase diagram, for three different directions of the
magnetic field \cite{TokiwaPhase}. In addition, several numerical investigations \cite{sheng,Sorella} of the
spatially anisotropic $J-J'$ Heisenberg model also find that in a wide region of the parameter space $0\leq
J^\prime/J \leq 0.6$ the spin chains are almost decoupled and inter-chain spin correlations are exponentially
weak. This behavior follows from highly frustrated structure of the inter-chain exchange $J^\prime$ and provides
additional justification to the 1D-based approach \cite{Starykh} to \ccc.

Electron spin resonance (ESR) represents a sensitive probe of low-energy magnetic excitations with zero momenta,
${\bf q} \approx 0$. It is well known as a method providing fine details of low-energy spin spectra in both
ordered  and paramagnetic \cite{esrbook,Ajiro,OshikawaAffleck} phases. Importantly, both the position and the
linewidth of the ESR signal are crucially sensitive to {\em anisotropies} of the spin Hamiltonian. Here we
report a dramatic manifestation of the {\em uniform} Dzyaloshinskii-Moriya (DM) interaction on the ESR spectrum
of \ccc\ in the paramagnetic phase. We observe and explain splitting of the ESR mode into two modes with
lowering the temperature, significantly extending previous study \cite{semeno}, and also find strong dependence
of the ESR absorption on the polarization of the microwave magnetic field ${\bf h}$ with respect to the DM
vector ${\bf D}$ in zero external field.

The crystal samples were grown by two methods. First set of samples was prepared by the crystallization from the
melt of a stoichiometric mixture of CsCuCl$_3$ and CsCl in an ampule, which was slowly pulled from the hot (T=
520$^\circ$C) area of the oven. The second method was crystallization from solution \cite{Soboleva}. The room
temperature crystal lattice parameters for both sets of crystals are a=9.75 {\AA}, b=7.607 {\AA}, c=12.394{\AA}.
The samples obtained by last method have natural faces and are lengthened along b-axes. Both sets of crystals
demonstrated identical ESR signals.
The study in the frequency range 9-90 GHz was performed with a set of homemade spectrometers equipped with
cryomagnets. The crystal samples were placed within microwave resonators at the maximum of the microwave
magnetic field. The resonance lines were taken at the fixed frequency, as dependences of microwave power,
transmitted through the resonator, on the external magnetic field. Temperatures down to 1.3 K (0.4 K), were obtained by
pumping vapor of $^4$He ($^3$He), correspondingly. Cooling down to 0.1 K was achieved with the help of a
dilution cryostat Kelvinox-400.

At temperatures above $T=10$ K crystals of Cs$_{2}$CuCl$_{4}$ exhibit conventional paramagnetic resonance of
Cu$^{2+}$ $S=1/2$ magnetic ions with a single narrow line corresponding to $g$-factor values
$g_{a}=2.20\pm0.02$, $g_{b}=2.08\pm0.02$, $g_{c}=2.30\pm0.02$ for the magnetic field applied  along ${\bf a}$, ${\bf b}$ and
${\bf c}$ axes correspondingly. These $g$-factor values agree well with those reported earlier
at $T=300$ K and $77$ K \cite{Sharnoff}, and, for $g_c$, at $T=4.2$ K \cite{semeno}.

\begin{figure}
    \includegraphics[scale=0.8]{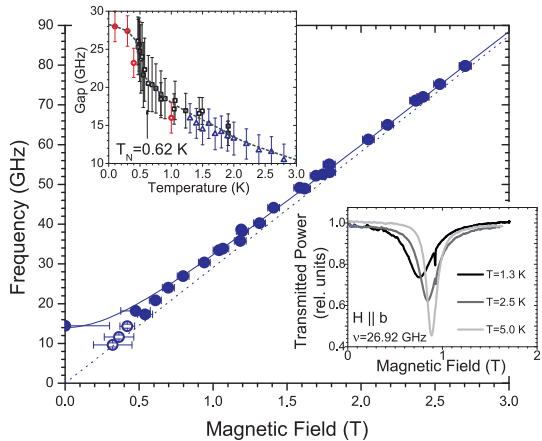}
    \caption{ESR frequency at $T=1.3$ K for $H \parallel {\bf b}$. Dotted line is paramagnetic resonance with $g=2.08$,
    solid line -- theory (see text).
Empty circles stand for resonances loosing intensity with cooling. Upper insert: gap vs temperature, various
symbols correspond to different cryostats. Lower insert:  evolution
    of the resonance line with cooling.\label{fHb}}
\end{figure}

\begin{figure}
    \includegraphics[scale=0.8]{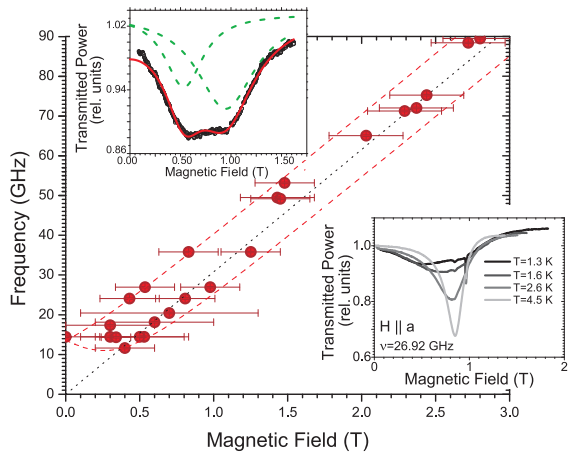}
    \caption{ESR frequencies at $T=1.3$ K for $H \parallel {\bf a}$. Dotted line is paramagnetic resonance with $g=2.20$,
    dashed line -- theory (see text). Upper
    insert: lineshape at $T=1.3$ K, $\nu=27$ GHz. 
    Points -- experimental data, solid line is the result of a fit by the sum of two Lorentzians (dashed lines).
    Lower insert: evolution of the resonance line with cooling.\label{fHa}}
\end{figure}

\begin{figure}
    \includegraphics[scale=0.8]{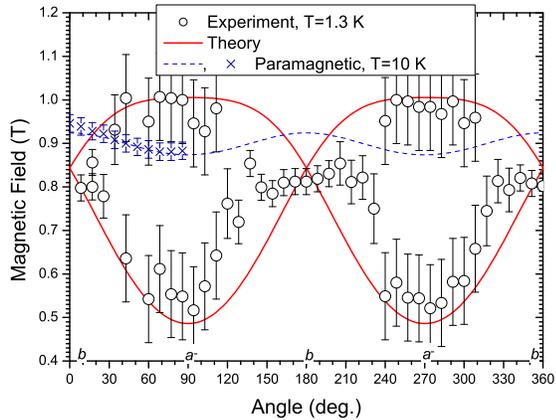}
    \caption{The angular dependence of the resonance field $H_{\rm r}$ in the ${\bf ab}$ plane for $\nu=27$ GHz at $T=1.3$ K.
    Thick line is the theoretical prediction with $D_{a}/(4\hbar)=8$ and $D_{c}/(4\hbar)=11$ GHz.
     The measured values of the resonance field deep in the paramagnetic phase (at $T =10$ K) are presented by crosses,
     the dashed line is a theoretical fit of a paramagnetic resonance with  $g_{a,b,c}$ of \ccc.
    \label{RotAB}}
\end{figure}

Significant evolution of the ESR spectrum was found on cooling below 6 K. This evolution depends strongly on the
microwave frequency $\nu$ and orientation of the external magnetic field ${\bf H}$.
({\it i}) For ${\bf H}\parallel {\bf b}$, we observe a single Lorentzian line shifting with
cooling to lower fields (see Fig. \ref{fHb}). At $\nu$=14 GHz the resonance field is near zero. At lower
frequencies, we observe a strong decrease of the intensity with cooling, and no visible shift from the
paramagnetic resonance field, in contrast to ESR lines taken at $\nu>$17 GHz, which exhibit a shift and do not
loose intensity  at cooling. At $\nu$=9 GHz the integral intensity at T=1.3 K is a half of that at T=4 K. The
resonant frequencies above 14 GHz may be well fitted by the frequency-field relation of an ordered
antiferromagnet with ``gap'' $\Delta/(2\pi\hbar)=14$ GHz,
\begin{equation}
   2\pi\hbar \nu=\sqrt{(g\mu_B H)^{2}+\Delta^{2}}.
    \label{gapped}
\end{equation}
({\it ii}) For ${\bf H} \parallel a, c$, the ESR line strongly broadens on cooling, and its shape distorts, as
shown in the inserts of Fig. \ref{fHa}. The absorption at $T$=1.3 K  may be fitted by a sum of two Lorentzians,
indicating the splitting of the resonance mode. The $\nu(H)$ dependence at ${\bf H}
\parallel a$ is presented in Fig. \ref{fHa}. Two modes at $T=1.3$ K were resolved for measurements in the
frequency range 14-50 GHz. At higher frequencies the splitting is
masked by natural linewidth, which prevents the resolution of the
doublet, while at lower frequencies the signal becomes too broad.
The angular $\phi$ dependence of the resonance field
${\bf H}_{\rm r}=H(\sin\phi, \cos\phi, 0)$ within the ${\bf a-b}$ plane, taken at
$\nu$=27 GHz and $T$=1.3 K, is shown in Fig.\ref{RotAB}. For
$\bf{H}\parallel c$ the splitting of about 0.35 $\pm0.1$ T is
resolved for frequencies 27 and 31 GHz.

\begin{figure}
    \includegraphics[scale=0.8]{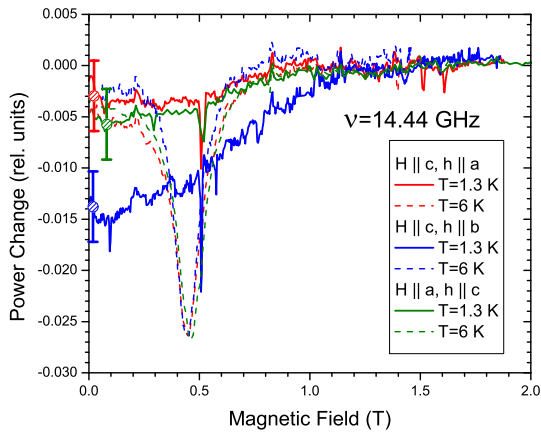}
    \caption{The polarization effect at $\nu=14.44$ GHz for $H\to 0$. Circles with error bars show the
 uncertainty resulting from normalizing the signals to zero absorption at $H > 1.5$ T. \label{Polarization14}}
\end{figure}

({\it iii}) We have also studied polarization dependence of the ESR absorption at $H\to 0$. Fig.~
\ref{Polarization14} shows transmission vs field dependences at $T=1.3$ K and $T=6$ K for different orientations
of the microwave and static fields with respect to the crystal axes, but with $\bf{h}\perp \bf{H}$. The
measurements are done at $\nu=14$ GHz. At $T>6$ K  both the position and the lineshape of the paramagnetic
resonance line are the same for all three principal polarizations of the microwave field. However, at $T=1.3$ K, the
absorption in zero field for $\bf{h}\parallel b$ is at least three times more intensive than for $\bf{h}
\parallel a,c$.

The splitting of the ESR line for ${\bf H} \parallel c$  and the polarization dependence of the signal were
also detected previously in Ref. \cite{semeno}.

We now turn to explanation of the data and show that all unusual
features of the ESR signal reported above are naturally explained
by quasi-1D nature of  \ccc. The relevant Hamiltonian
\begin{eqnarray}
{\cal H} &=& \sum_{x,y,z} J {\bf S}_{x, y, z} \cdot {\bf S}_{x+1,
y, z} - {\bf D}_{y,z} \cdot {\bf S}_{x, y, z} \times {\bf S}_{x+1,
y, z} +
\nonumber\\
&& - g \mu_B {\bf H} \cdot {\bf S}_{x,y,z} + ... \label{eq:H}
\end{eqnarray}
contains three terms: the first describes intrachain exchange $J$ ($x$ runs along crystal ${\bf b}$ axis), the
second -- uniform DM interaction ${\bf D}_{y,z}$ between chain spins, and the third is the usual Zeeman term,
allowing for anisotropic $g$-factor. The dots stand for the omitted interchain exchange as well as DM
interactions on interchain bonds. Detailed symmetry analysis of the allowed DM interactions \cite{Starykh} shows
that there are {\em four} different orientations of the DM vector (see Fig. 6 in \cite{Starykh}) depending on
chain's integer coordinates $y, z$: ${\bf D}_{y,z} = D_a (-1)^z \hat{a} + D_c (-1)^y \hat{c}$. Here $z$ indices
magnetic ${\bf b-c}$ layers, while $y$ numerates chains within a layer. Crucially, crystal symmetry forbids DM
vector to have component along the chain ${\bf b}$ axis.\cite{Starykh}

\begin{figure}
    \includegraphics[scale=0.5]{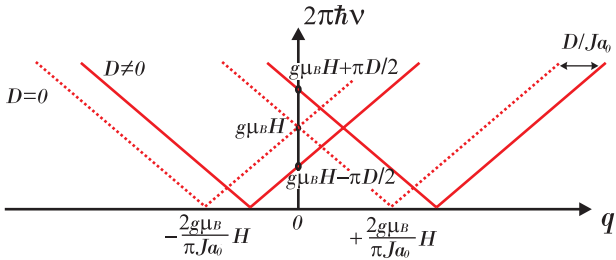}
    \caption{Spinon spectrum of $S=1/2$ Heisenberg chain for ${\bf q} \sim 0$ with (solid lines) and without (dashed lines)
    DM interaction. When ${\bf D} \parallel {\bf H}$, momentum is boosted by $q=D/Ja_0$.
 \label{spectrum}}
\end{figure}

The essence of the observed ESR line splitting can be explained by considering a single Heisenberg chain with
uniform DM interaction with vector $\bf{D}$ along the local $\hat{z}$ axis, when the field is lined up along the
same axis (${\bf D}_{y,z} \to D \hat{z}$, ${\bf H} \to H \hat{z}$). In this geometry the DM interaction can be
gauged away by position-dependent rotation of  {\em lattice} spins,
\begin{equation}
S^+_{x,y,z} = \tilde{S}^+_{x,y,z} e^{i \alpha x}, ~S^z_{x,y,z} =\tilde{S}^z_{x,y,z}.
\label{eq:boost}
\end{equation}
The rotation angle $\tan(\alpha) \approx \alpha = - D/J$ is chosen so as to eliminate the DM coupling from the Hamiltonian. The
transformed $\tilde{{\cal H}}$ describes spin chain with {\em quadratic} in $D$ easy-plane anisotropy. To
linear in $D/J$ accuracy the ESR response of this model coincides with that of an isotropic Heisenberg chain,
detailed study of which is described in Ref.\cite{OshikawaAffleck}. ESR absorption is determined by the transverse
structure factor ${\cal S}_{xx}(\omega, \tilde{q})$ = ${\cal S}_{yy}(\omega, \tilde{q})$ of the chain, evaluated
at $\tilde{q}=0$. This is produced by characteristic to the spin-1/2 chain {\em spinon} continuum and is described by a sum of
delta-function peaks at frequencies $\hbar \omega_\pm = |\hbar v \tilde{q} \pm g\mu_B H|$, see eq.(3.16) of
\cite{OshikawaAffleck} and Fig.\ref{spectrum}. Here $v = \pi J a_0/(2 \hbar)$ is the zero-field spinon velocity
and $a_0$ is the lattice spacing \cite{support}.
The response in the {\em original} (un-rotated) basis is obtained
by un-doing the momentum boost \eqref{eq:boost}, which corresponds to setting $\tilde{q} = D/J a_0$ in the
expression for the structure factor ${\cal S}_{xx/yy}$. This immediately implies the splitting of the ESR line
into two lines, at frequencies $\hbar\omega_\pm = |g\mu_B H \pm \pi D/2|$. Note that for small $D/H$ the
splitting $\pi D = \pi (D_a^2 + D_c^2)^{1/2}$ is linear in $D$ which justifies our neglect of the easy-plane anisotropy.

While the four-sublattice structure of the DM vectors makes it impossible to study the ${\bf D} \parallel {\bf H}$
configuration in \ccc , the above argument allows one to immediately understand polarization-dependent
absorption (finding ({\it iii}) above) in zero magnetic field, at the frequency $2\pi \nu = \pi D/2\hbar$. Since
the microwave absorption is proportional to the square of the microwave field component perpendicular to the
effective field, we conclude that configuration with ${\bf h}\perp {\bf D}$ (that is, ${\bf h} \parallel {\bf b}$)
should result in maximal possible absorption. For ${\bf h}$ along the ${\bf a}$ (${\bf c}$) axis, absorption is
a factor $D_c^2/D^2$ ($D_a^2/D^2$) smaller. Using numerical estimates of $D_{a,c}$ derived below, the absorption
for different polarizations should obey the relations: $P_{b}/P_{a}\simeq 1.6$ and $P_{b}/P_{c}\simeq 2.8$. This
agrees qualitatively with our data presented in Fig.~\ref{Polarization14} -- the zero field absorption at
$\nu=14$ and $17$ GHz is a factor of 3 more intensive for ${\bf h} \parallel {\bf b}$
than for ${\bf h} \parallel {\bf a}, {\bf c}$ .

To understand our findings ({\it i}) and ({\it ii}), one needs to
analyze the arbitrary orientation of the ${\bf H}$ and ${\bf D}$
vectors, which is the problem solved in \cite{Starykhchains}.
The main physical point of Ref.\cite{Starykhchains} consists in the observation that a uniform DM interaction,
much like a spin-orbit Rashba interaction in 2D conductors \cite{kalevich,jantsch,folk}, acts on spinon
excitations of 1D chains as an internal momentum-dependent magnetic field.
Technically, this observation follows from spin-current formulation \cite{support} of the problem \eqref{eq:H},
which is valid below the strong-coupling temperature scale $T_0
\sim J e^{-\pi S}$ ($S=1/2$ for the spin chain here)\cite{buragohain}.
The spin currents ${\bf M}_{R/L}$ represent spin density fluctuations of spinons
near the right/left (R/L) Fermi points of a 1D system.
It follows that the right and left-moving currents experience {\em different},
in magnitude and direction, {\em total} magnetic fields: ${\bf B}_R = {\bf H} + \hbar v {\bf D}/g \mu_B J$
and ${\bf B}_L = {\bf H} - \hbar v {\bf D}/g \mu_B J$. It is then natural that the ESR response of such a chain
consists of two peaks at frequencies $2\pi \hbar \nu_{\rm R/L} = g \mu_B B_{R/L}$.\cite{Starykhchains}

For the \ccc-specific configuration of chain DM interactions and
${\bf H} = (H_a, H_b, H_c)$ our analysis predicts
\begin{eqnarray}
 (2 \pi \hbar \nu_{\rm R})^2  &=&    (g_{b}\mu_{B}H_{b})^{2} +  (g_{a}\mu_{B}H_a + (-1)^z \pi D_{a}/2)^{2}+  \nonumber\\
 &&(g_{c}\mu_{B}H_c  + (-1)^y \pi D_{c}/2)^{2},
\label{freqStarykh1}
\\
(2 \pi \hbar \nu_{\rm L})^2  &=&   (g_{b}\mu_{B}H_{b})^{2} +  (g_{a}\mu_{B}H_a - (-1)^z \pi D_{a}/2)^{2}+   \nonumber\\
 &&(g_{c}\mu_{B}H_c - (-1)^y \pi D_{c}/2)^{2},
    \label{freqStarykh2}
\end{eqnarray}
Observe that while $\nu_R \neq \nu_L$ for fixed chain indices $(y,z)$, these {\em chiral} ESR frequencies transform into each
other under lattice translations $(y,z) \to (y\pm 1, z\pm 1)$.
These equations naturally explain the difference between
${\bf H} \parallel {\bf a}, {\bf c}$ and ${\bf H}\parallel {\bf b}$ situations.
For ${\bf H}\parallel {\bf b}$ the external and the internal DM field are mutually perpendicular and
the vector sum ${\bf H \pm D}$ has the the same absolute value for both signs.
One then finds single ESR frequency $\nu_R = \nu_L$ of the
form \eqref{gapped} where the gap is in fact determined by DM interaction, $\Delta=\pi \sqrt{D_{a}^{2}+D_{c}^{2}}/2$.

For ${\bf H} \parallel {\bf a}, {\bf c}$  the DM field has a component along ${\bf H}$ and the frequencies of
the two spin excitations are different. In principle, because of the four values of the ${\bf D}_{y,z}$ vector,
four different frequencies may be present when the field ${\bf H}$ is oriented within the ${\bf a-c}$ plane. We
attempted to check this experimentally but failed to resolve separate lines within the broad band of
absorption.\cite{support} For ${\bf H}$ within the ${\bf a-b}$ plane \eqref{freqStarykh1} and
\eqref{freqStarykh2} predict two different frequencies.  Solving equations
(\ref{freqStarykh1},\ref{freqStarykh2}) at the fixed frequency for the magnetic field we were able to fit the
experimental angular dependence in Fig. \ref{RotAB}. In this way we obtained $D_a/(4\hbar)=8 \pm 2$ and
$D_c/(4\hbar)=11 \pm 2$ GHz, which corresponds to $0.29\pm0.07$ and $0.39\pm0.07$ T. We have to note here that
for ${\bf H} \parallel {\bf c}$ the observed splitting of the frequency doublet $\Delta \nu = 0.35$ T is about
$50\%$ smaller than the prediction of (\ref{freqStarykh1},\ref{freqStarykh2}) with these $D_{a/c}$ values.This
deviation from the predicted value is dicussed in \cite{support}.

A more detailed comparison between the experiment and the theory would require extending the latter to finite
temperatures since the reported measurements are performed in the intermediate temperature range $T_{\rm N} =
0.62 {\text{K}} < T <T_{\rm CW} = 4 {\text{K}}$. Temperature dependence of the gap $\Delta$ for ${\bf
H}\parallel {\bf b}$ in a wider range including $T_N$ is shown on Fig.\ref{fHb}. The variation of the gap in the
range 1~K$<T<$2~K is much slower than right near $T_{\rm N}$. This  supports our main assumption that the
reported spectral features of \ccc\ reflect specific spin chain physics which dominates the temperature range
$T_{\rm N} < T < T_{\rm CW}$.

In conclusion, we demonstrated that the uniform DM interaction, which is a distinctive feature of
Cs$_{2}$CuCl$_{4}$, results in a new kind of spin resonance in a S=1/2 chain antiferromagnet. The observed
spectrum is a consequence of the splitting of the chain's spinon continuum by the internal magnetic field
which is produced by the uniform DM interaction. We believe that a similar phenomenon is
possible in a higher-dimensional magnetic systems with fractionalized spinon excitations.
Our findings differ strongly from the well-known single-frequency resonance of  $S=1/2$ antiferromagnetic spin chain with
staggered DM interaction \cite{OshikawaAffleck}. It also differs from a conventional 3D antiferromagnet which
acquires an energy gap in the spectrum only below the ordering transition.

We would like to thank K.O. Keshishev for  guidance in work with the dilution refrigerator, and L. Balents, S.
Gangadharaiah, V.N. Glazkov, M. Oshikawa, and R. Moessner for insightful discussions. We thank RFBR Grant
09-02-736 and NSF Grant No. DMR-0808842 (O.A.S.) for financial support. Part of the work has been performed
within the ASG on ``Unconventional Magnetism in High Fields'' at the MPIPKS (Dresden).

\onecolumngrid
\newpage
\input{povarov2}

\end{document}

%% file: povarov2.tex
\setcounter{figure}{0}
\setcounter{equation}{0}

\twocolumngrid

{\bf{Supporting material for \\
 ``Modes of magnetic resonance in the spin liquid phase of \ccc.''}}

\begin{figure}[h]
    \includegraphics[scale=0.4]{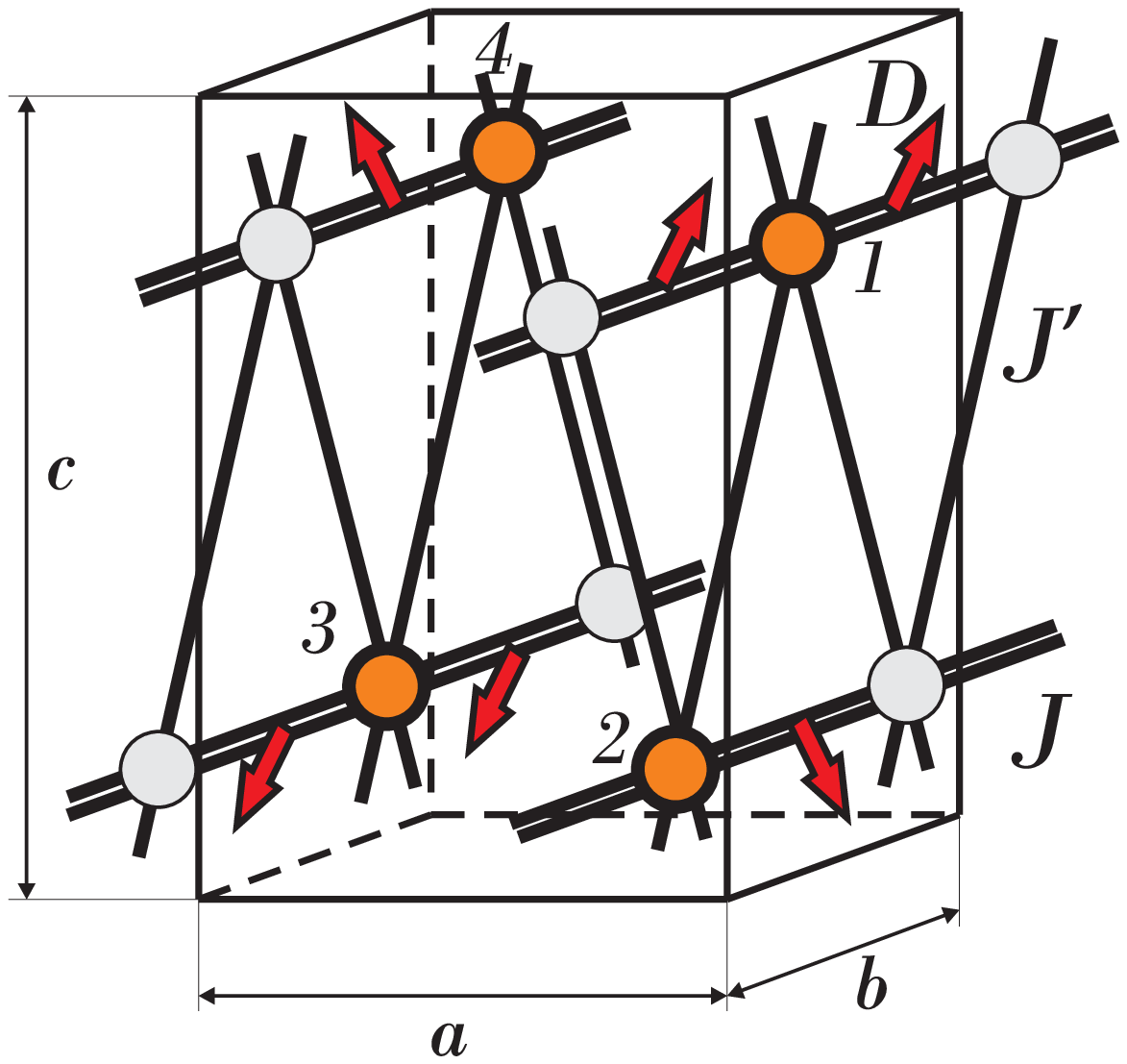}
    \caption{Parameters of spin interactions in \ccc.  Four different orientations of the chain DM vector
    ${\bf D}$ are illustrated by red arrows.}
    \label{Structure}
\end{figure}

In this work we report results of electron spin resonance (ESR) study of quasi two-dimensional (2D)
antiferromagnet \ccc\ in the paramagnetic phase. \ccc\ realizes distorted triangular lattice structure, shown in
Fig.\ref{Structure}, with  the strong exchange $J$ along  the base of the triangular cell and with the weaker
exchange integral $J^\prime$ for the exchange paths along lateral sides of the triangle, their ratio is $J'/J
=0.34$. In-chain DM interaction $${\bf D}_{y,z} = D_a (-1)^z \hat{a} + D_c (-1)^y \hat{c},$$ derived in Ref.
\onlinecite{Starykh2}, describes four possible orientations of the ${\bf D}$ vector on the chain with integer
indices $y,z$, see Figure \ref{Structure}. Here $y$ and $z$ axes are aligned along $a$ and $c$ crystallographic
axis of an orthorombic Pnma lattice, respectively.

\section{Angular dependences of the resonance fields}

\begin{figure}
    \includegraphics[scale=0.8]{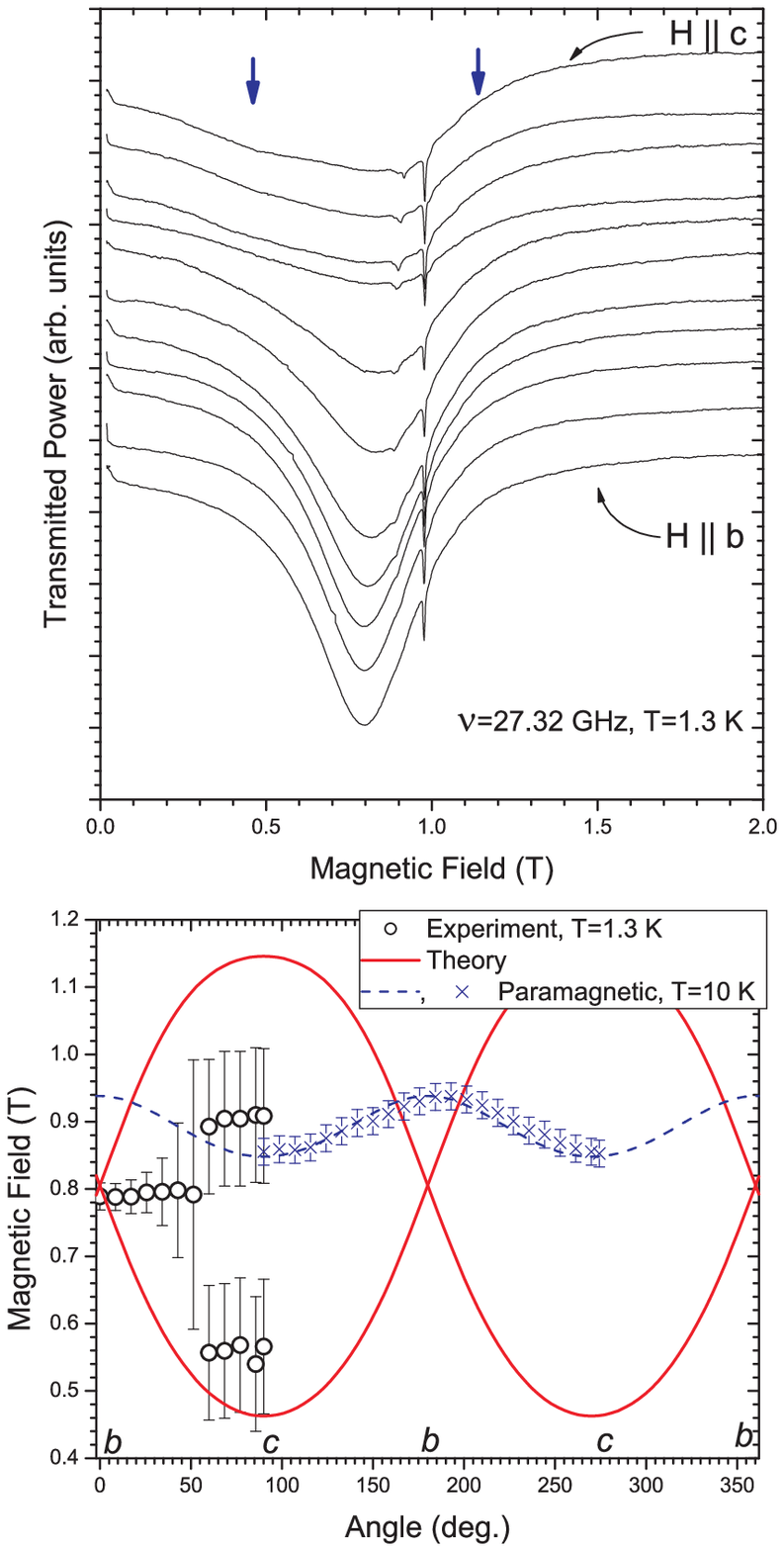}
 \caption{Upper panel: $\nu =27$ GHz resonance lines
for the magnetic field ${\bf H}$ in the ${\bf bc}$-plane. The curves are taken at the equidistant angles. Blue
arrows for ${\bf H}\parallel {\bf c}$ curve indicate the calculated resonance fields for the splitted ESR line.
Lower panel: The $\psi$ angular dependence of the resonance field in ${\bf bc}$-plane for $\nu=27$ GHz at
$T=1.3$ K. Thick line is the theoretical  calculation (see text) with $D_a/(4\hbar)=8$ and $D_c/(4\hbar)=11$
GHz. Dashed line shows calculated angular dependence for a paramagnet with g-factors of \ccc, crosses
represent experimental values at $T = 10$ K (deep in the paramagnetic phase).} \label{Rotbc}
\end{figure}

In addition to  the angular dependence of the resonance field  on the angle $\phi$ within the ${\bf ab}$-plane,
presented in the main text, we present in Fig.\ref{Rotbc} the angular dependence of the resonance field on the
angle $\psi$ within the the ${\bf bc}$-plane. In this case the agreement is only qualitative: the splitting at
$\bf{H} \parallel c$, obtained by a two-Lorentzian fit constitutes about 50\% of the predicted by the DM-theory.
Nonetheless, the predicted splitting of the resonance fields is within the broad band of absorption, observed at
$\bf{H} \parallel {\bf c}$, as indicated by blue arrows in the upper panel of Fig. \ref{Rotbc}.

\begin{figure}
  \includegraphics[scale=0.8]{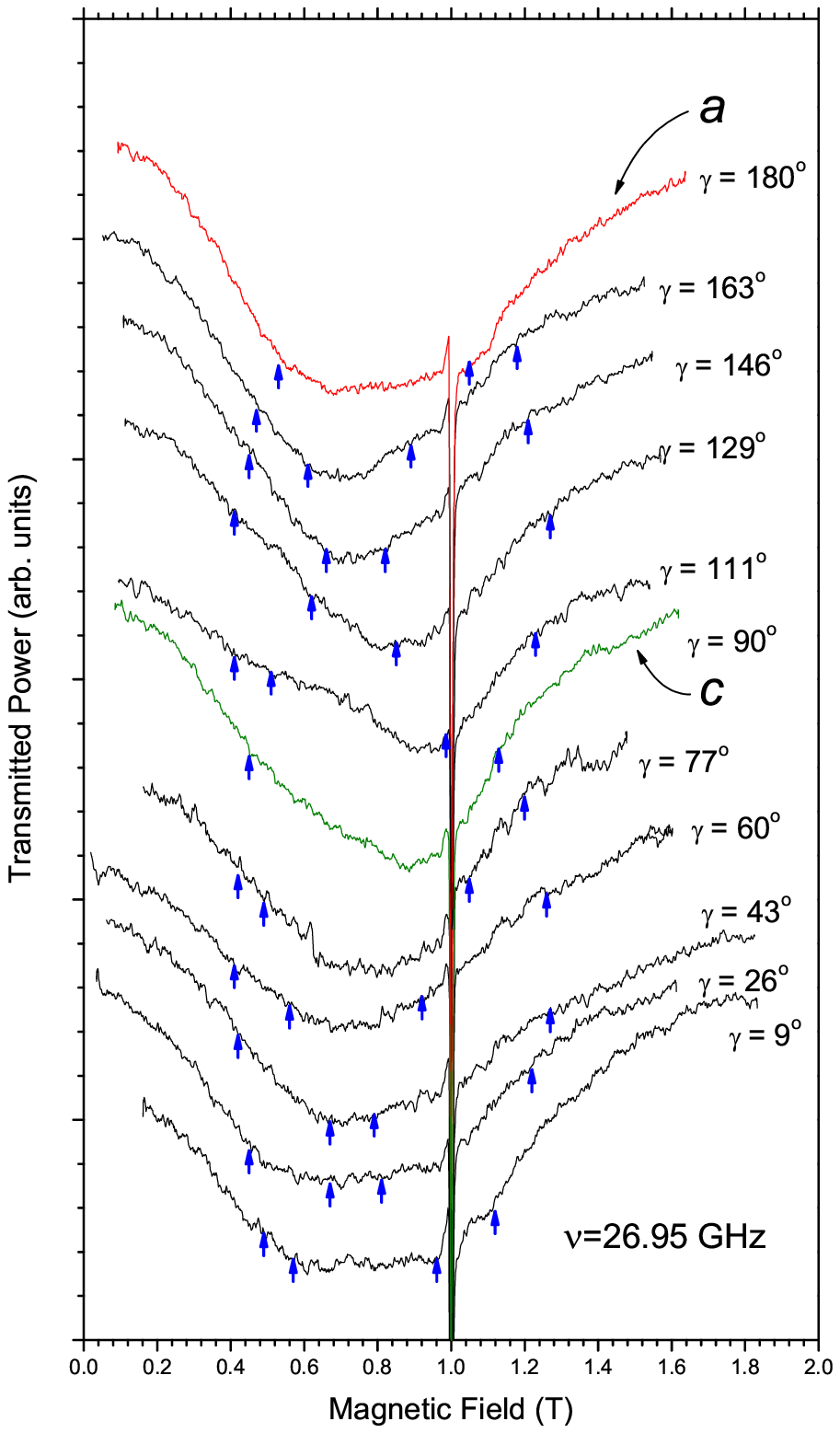}
    \caption{The ESR lines for $\nu=27$ GHz and $T=1.3$ K at different orientations of the static field
     within the ${\bf ac}$-plane. Arrows mark the
resonances, expected at $T=0$ for decoupled  spin chains according
to the DM theory (see main text) with
     $D_{a}/4\hbar=8$ and $D_{c}/4\hbar=11$ GHz.}
        \label{Rotac}
\end{figure}

\begin{figure}
    \includegraphics[scale=0.8]{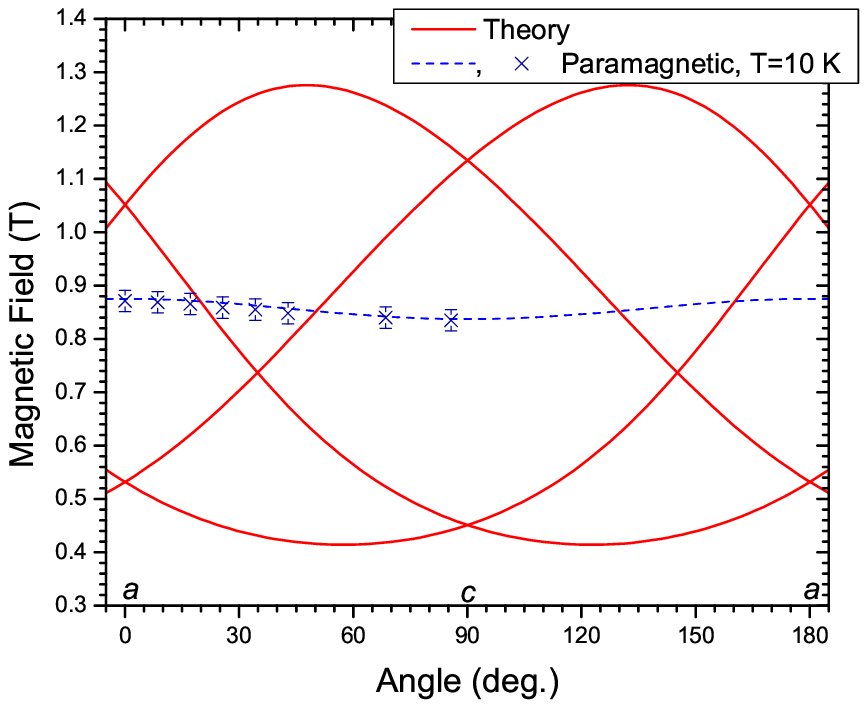}
    \caption{Resonance fields  for  $\nu = 27$ GHz ESR as a function of the angle
    in the ${\bf ac}$-plane according to the DM-theory with
     $D_{a}/4\hbar=8$ and $D_{c}/4\hbar=11$ GHz. Dashed line shows calculated angular dependence for
      a paramagnet with g-factors of \ccc, crosses
represent experimental values at $T = 10$ K (deep in the paramagnetic phase)
        \label{Rotac_theory}}
\end{figure}

Fig.\ref{Rotac} presents the evolution of the absorption band for the field oriented within the ${\bf ac}$-plane.
According to our theory (see main text) the ESR line should split into four resonances, see Fig.\ref{Rotac_theory}.
We failed to resolve four lines because the width of the individual lines is too broad.
Nevertheless, the expected positions of four lines, marked by arrows in Fig.\ref{Rotac},
do lie within the measured absorption bands.

\section{Frequency-field  and temperature dependences for ${\bf H} \parallel {\bf c}$}

\begin{figure}
   \includegraphics[scale=0.8]{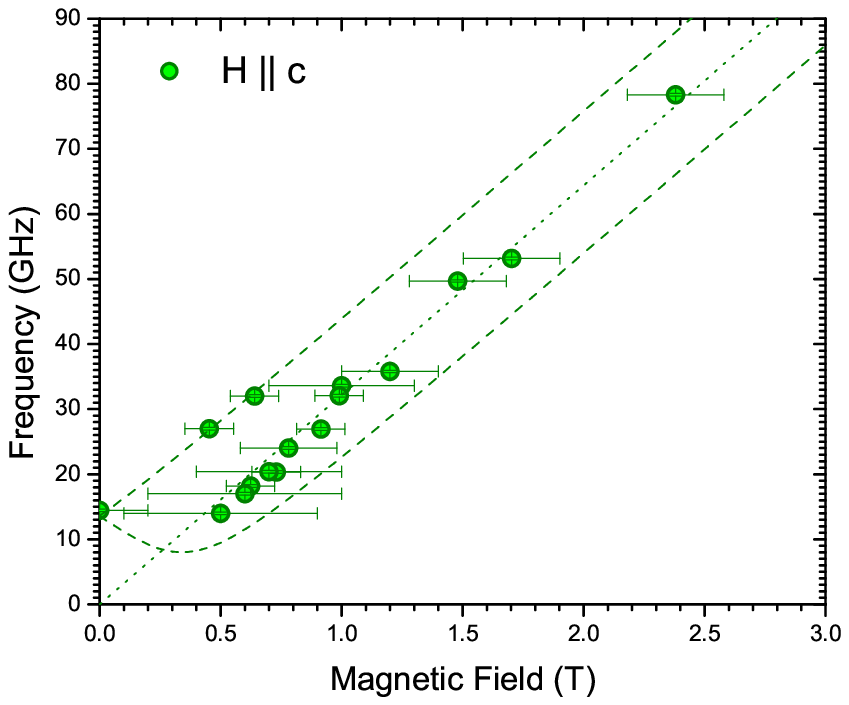}
    \caption{ESR frequencies at $T=1.3$ K for $H \parallel {\bf c}$. Dotted line is paramagnetic resonance with $g=2.30$,
    dashed line -- DM theory (see text), $D_{a}/4\hbar=8$ and $D_{c}/4\hbar=11$ GHz.
    \label{fHc}}
\end{figure}

\begin{figure}
\includegraphics[scale=0.8]{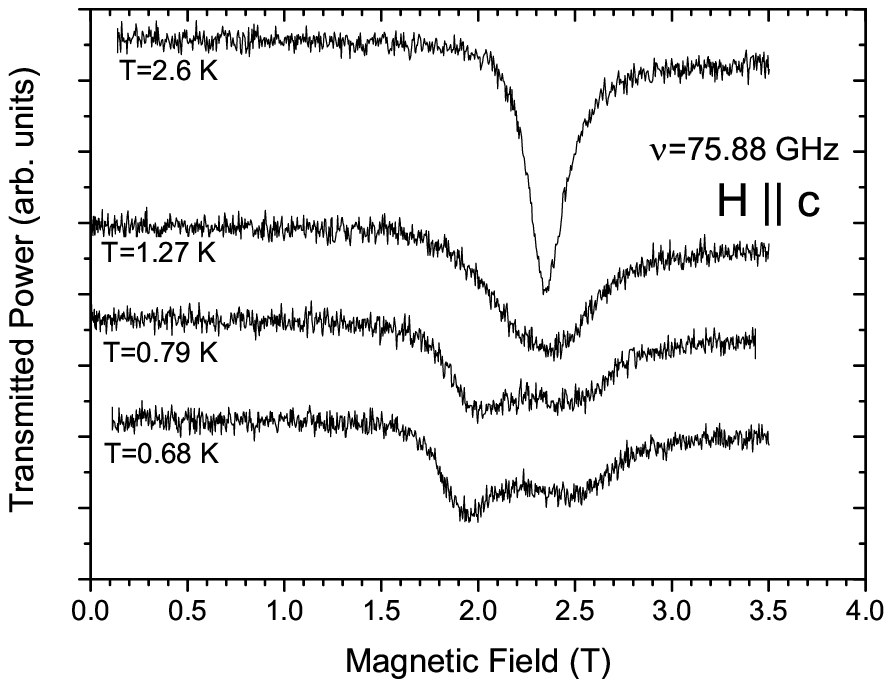}
    \caption{$\nu = 75.88$ GHz ESR lines at different temperatures for $H \parallel {\bf c}$.\label{LTHc}}
\end{figure}

The frequency-field dependence for ${\bf H} \parallel {\bf c}$ at $T=1.3 K$ is shown in the Fig.\ref{fHc}. At $T=1.3$
K the splitting is resolved only for two frequencies in the middle of the frequency range. At low frequencies,
the large linewidth masks the splitting, while at higher frequencies the splitting predicted by the DM-theory is
not observed at this temperature. However, at lower temperatures the splitting of the ESR line for ${\bf H} \parallel {\bf c}$
becomes more visible. It is clearly seen even at the high frequency edge of the range, $\nu = 76$ GHz,
as is seen in Fig. \ref{LTHc}.  The splitting observed at $T=0.79$ K equals $0.6\pm 0.1$ T,
which corresponds satisfactory to the predicted value of $0.75$ T
(horizontal shift between the asymptotic parts of dashed lines in Fig.\ref{fHc}).
For ${\bf H} \parallel {\bf a}$ and at $T=0.8$ K, we observe similar
splitting of $0.6\pm 0.1$ T for the frequency $\nu =72.9$ GHz.

\section{Temperature dependence of ESR lines at frequencies above and
below the gap $\Delta$ for ${\bf H} \parallel {\bf b}$}
\label{sec:freq-T}

It was mentioned in the main text that  at the frequencies below the DM gap, i.e. $\nu < \Delta =14$ GHz, the
ESR line does not exhibit a shift but instead looses the intensity. This observation is illustrated in Fig.
\ref{b927Ghz}, where the data for $\nu =9.63$ GHz ESR signal are presented. One can clearly see that the
resonance field does not change with lowering the temperature, while the intensity of the line does diminish
strongly.

To contrast this with quite different evolution of the signal at higher frequencies $\nu > \Delta$, we present
in Fig.\ref{b927Ghz} detailed temperature evolution of the ESR line at $\nu= 26.9$  GHz. Here the shift of the
line's center to the lower values of the resonance magnetic field is obvious. The signal also broadens such that
the overall intensity is conserved.

For other field orientations, in particular for ${\bf H} \parallel {\bf a,c}$,
the ESR line completely disappears for low frequencies $\nu < \Delta$.

\section{Theoretical details}
\subsection{Lattice Hamiltonian for ${\bf D}\parallel {\bf H}$}

When ${\bf H} \parallel {\bf D}$, unitary rotation (3) of the paper transforms chain Hamiltonian (2) into
\begin{eqnarray}
{\tilde{\cal H}} &=& \sum_{x} \{\sqrt{J^2 + D^2}  [{\tilde S}^x_{x} {\tilde S}^x_{x+1} +
{\tilde S}^y_{x} {\tilde S}^y_{x+1}] + J {\tilde S}^z_{x} {\tilde S}^z_{x+1}
\nonumber\\
&& - g \mu_B H {\tilde S}^z_{x},
\label{eq:H-tilde}
\end{eqnarray}
where for brevity we do not write chain's integer coordinates $y,z$. As claimed in the paper the obtained
Hamiltonian \eqref{eq:H-tilde} possesses an easy-plane anisotropy of the strength $D^2/2 J$. Neglecting it takes
us to the Heisenberg limit discussed in the paper and illustrated in Fig.5 there.

\subsection{Low-energy theory}

The obtained Hamiltonian \eqref{eq:H-tilde} can also be used to derive the low-energy form of the DM coupling.
This is done by bosonizing \eqref{eq:H-tilde} (with $H=0$) which results in \cite{gogolin_book2} \be
{\tilde{\cal H}} = \int dx ~\frac{v}{2}\{ (\partial_x \tilde \varphi)^2 + (\partial_x \tilde \theta)^2\},
\label{eq:H-bos} \ee where $(\tilde \varphi, \tilde \theta)$ is the pair of conjugated bosonic fields which
describes low-energy excitations in the {\em rotated} basis. In particular, the transverse component of the spin
density operator (at momentum $\pi$) is expressed via $\tilde \theta$ as \cite{gogolin_book2} \be \tilde{S}^+_x
= A_3 e^{i\beta \tilde{\theta}} \label{eq:S+} \ee where $\beta=2\pi R$ determines scaling dimension of the spin
operator in terms of the known compactification radius $R$, see Ref.\onlinecite{oa1999-2}, and $A_3$ is
non-universal amplitude. Also note that spinon velocity $v$ in \eqref{eq:H-bos} is function of magnetic field.
It monotonically decreases with $H$ and reaches zero at the saturation \cite{oa1999-2}. Near $H=0$ the $v(H)$
dependence is weak and we can safely approximate $v$ by its zero-field value everywhere in this study. Thus,
when necessary, we use \be v = \frac{\pi}{2} \frac{J a_0}{\hbar} \label{eq:v} \ee for the velocity $v$. Here
$a_0$ is the lattice constant in units of which we measure coordinate $x$.

Comparing \eqref{eq:S+} with equation (3) of the paper we immediately observe
that at low energies that unitary rotation simply corresponds to the shift of momentum
by $\alpha=-D/J$ (which is exactly the physics discussed in the main paper).
Hence  bosonic field $\theta$ of the original theory and $\tilde{\theta}$ of the
rotated one are simply related as
\be
\theta = \tilde{\theta} - \frac{D}{\beta J} x .
\label{eq:theta}
\ee
Hence the low-energy theory of the original Heisenberg chain with uniform DM
interaction can be obtained from \eqref{eq:H-bos} via \eqref{eq:theta} so that
\bea
{\cal H} &=& \int dx ~\frac{v}{2}\{ (\partial_x \varphi)^2 + (\partial_x \theta + \frac{D}{\beta J})^2\} =
\nonumber\\
&=&\int dx ~\frac{v}{2}\{ (\partial_x \varphi)^2 + (\partial_x \theta)^2\}  + \frac{v D}{\beta J} \partial_x
\theta + \text{const}. \label{eq:H-DM} \eea Using that at $H=0$ $\beta=\sqrt{2\pi}$ and $M^z_R - M^z_L =
-\partial_x \theta/\beta$ we find that the last term in \eqref{eq:H-DM} is proportional to the difference of the
right and left spin currents. This allows us to identify the lattice and the low-energy forms of the DM
interaction as \be \sum_x D{\hat z}\cdot {\bf S}_x \times {\bf S}_{x+1} \to -\frac{vD}{J}\int dx ~(M^z_R -
M^z_L) \label{eq:DM} \ee This result is written in terms of fluctuating spin densities (spin currents
\cite{gogolin_book2}) $M^a_{R/L}$ defined in terms of (Dirac) fermions $\Psi_{R/L}$ near the right/left (R/L)
Fermi points, correspondingly. More precisely, \be M^a_{R/L} = \Psi_{R/L;s}^\dagger \frac{\sigma^a_{ss'}}{2}
\Psi_{R/L;s'}. \label{eq:M} \ee It turns out that the spin-current formulation is particularly convenient for
the problem at hand (see \cite{osnano2} for pedagogical explanation of the technique).

To complete the low-energy description we also need the low-energy form of the Zeeman coupling,
which is well-known and reads (for the moment we keep magnetic field along $\hat{z}$ axis as well)
\bea
&&\sum_x g\mu_B H S^z_x \to \frac{g\mu_B H}{\sqrt{2\pi}}\int dx ~\partial_x \varphi = \nonumber\\
&& = g\mu_B H \int dx ~(M^z_R + M^z_L). \label{eq:zeeman} \eea In writing \eqref{eq:H-bos} we have omitted a
number of marginal terms which account for the easy-plane anisotropy as well as residual backscattering
interaction between right- and left- spin-currents \cite{Starykhchains2,garate2}. These terms do not affect our
main result - the splitting of the ESR signal into two due to the interplay of the external ${\bf H}$ and the
internal DM ${\bf D}$ fields. They are, most likely, important for understanding the {\em width} of the
individual ESR line, which is the problem of much greater theoretical complexity and is left for future studies.

With the help of \eqref{eq:DM} and \eqref{eq:zeeman} we can now write down the low-energy Hamiltonian
of the Heisenberg chain subject to both Zeeman and DM interactions acting along different directions,
\bea
{\cal H}_{\rm chain} &=& {\cal H}_0 - \int dx~ g\mu_B H (M^h_R + M^h_L) +\nonumber\\
&&-\int dx ~(v D/J)(M^d_R - M^d_L) \label{eq:Hz+dm} \eea where Hamiltonian ${\cal H}_0$ of the ideal Heisenberg
chain is written in Sugawara form \cite{gogolin_book2} as \be {\cal H}_0 = \frac{2\pi v}{3} \int dx ~\{{\bf M}_R
\cdot {\bf M}_R + {\bf M}_L \cdot {\bf M}_L \} \label{eq:H0} \ee To connect with previous expressions, it is
worth noting that \eqref{eq:H0} is in fact equivalent to the abelian bosonization form \eqref{eq:H-bos}. The
more complicated-looking \eqref{eq:H0} has one significant advantage over \eqref{eq:H-bos}: it makes evident a
remarkable emerging SU(2)$_R\times$ SU(2)$_L$ symmetry \cite{gogolin_book2}, which allows for independent
rotations of the right- and left- spin currents. This key for the following symmetry emerges below the
strong-coupling energy/temperature scale $T_0 \sim J e^{-\pi S}$ ($S=1/2$ for the spin chain
here)\cite{buragohain2}, when the spin current formulation becomes appropriate. Above that temperature a more
conventional semi-classical description of the spin-1/2 chain in terms of fluctuating N\'eel vector order
parameter field is valid \cite{buragohain2}. In fact, one can think of $T_0$ scale as of `Haldane' temperature
below which the difference between {\em integer} and {\em half-integer} spin chains becomes pronounced. It is
the temperature (or, equivalently, energy) scale below which the true {\em spinon} nature of the quantum ground
state of the critical spin-1/2 chain is manifest.

Note that upper indices $d$ and $h$ on currents ${\bf M}_{R/L}$ in \eqref{eq:Hz+dm}
denote directions of the vectors ${\bf D}$ and ${\bf H}$.
Observe that DM coupling (last term in ${\cal H}_{\rm chain}$) is odd
under spatial inversion which interchanges $R$ and $L$ -- this symmetry distinction is already evident
in the lattice Hamiltonian (2) of the main paper, see also \eqref{eq:DM}.

\subsection{ESR as a chiral probe}

The remaining steps closely follows Ref. \onlinecite{Starykhchains2}. We now perform independent rotations of
the right- and left- currents in \eqref{eq:Hz+dm}, so that in the rotated basis \be {\cal H}_{\rm chain} = {\cal
H}_0 - g\mu_B \int dx (B_R M^z_R + B_L M^z_L) \label{eq:Hrot} \ee the Hamiltonian describes unusual situation
whereby right and left-moving currents experience {\em different}, both in magnitude and direction, magnetic
fields: ${\bf B}_R = {\bf H} + \hbar v {\bf D}/g \mu_B J$ and ${\bf B}_L = {\bf H} - \hbar v {\bf D}/g \mu_B J$.
Observe that such a rotation does not affect ${\cal H}_0$ which retains its quadratic in spin currents form
\eqref{eq:H0}.

It is now rather natural (see Refs. \onlinecite{Starykhchains2,demartino2} for details) that the ESR response
the spin chain consists of two peaks at frequencies (here we make use of \eqref{eq:v}) \be 2\pi \hbar ~\nu_{\rm
R/L} = g \mu_B |{\bf B}_{R/L}| = |g \mu_B {\bf H} \pm \pi {\bf D}/2|. \label{eq:nu2} \ee The obtained result
includes ${\bf H}\parallel {\bf D}$ and ${\bf H}=0$ situations discusses in the main paper as special cases.

It is interesting to note that in the described case of a single spin chain with constant ${\bf D}$ vector
the ESR experiment becomes {\em chiral} probe: \eqref{eq:nu2} shows that right- and left- moving
excitations can be accessed independently by simply varying (and rotating) external magnetic
field ${\bf H}$ with respect to DM axis ${\bf D}$. In the case of \ccc\ this interesting point is lost
as the DM axis takes on four different directions, depending on the $y,z$ coordinates of the spin chain.
As a result the {\em right} ESR frequency, equation (4) of the main paper, for, say, chain with {\em even}
coordinates $y,z$ coincides with the {\em left} frequency (equation (5)) from chains with {\em odd} $y,z$ indices.
Thus, the crystal structure of the material masks the chiral nature of the ESR probe.

\subsection{Frequency dependence of the ESR response}

\begin{figure}
    \includegraphics[scale=0.8]{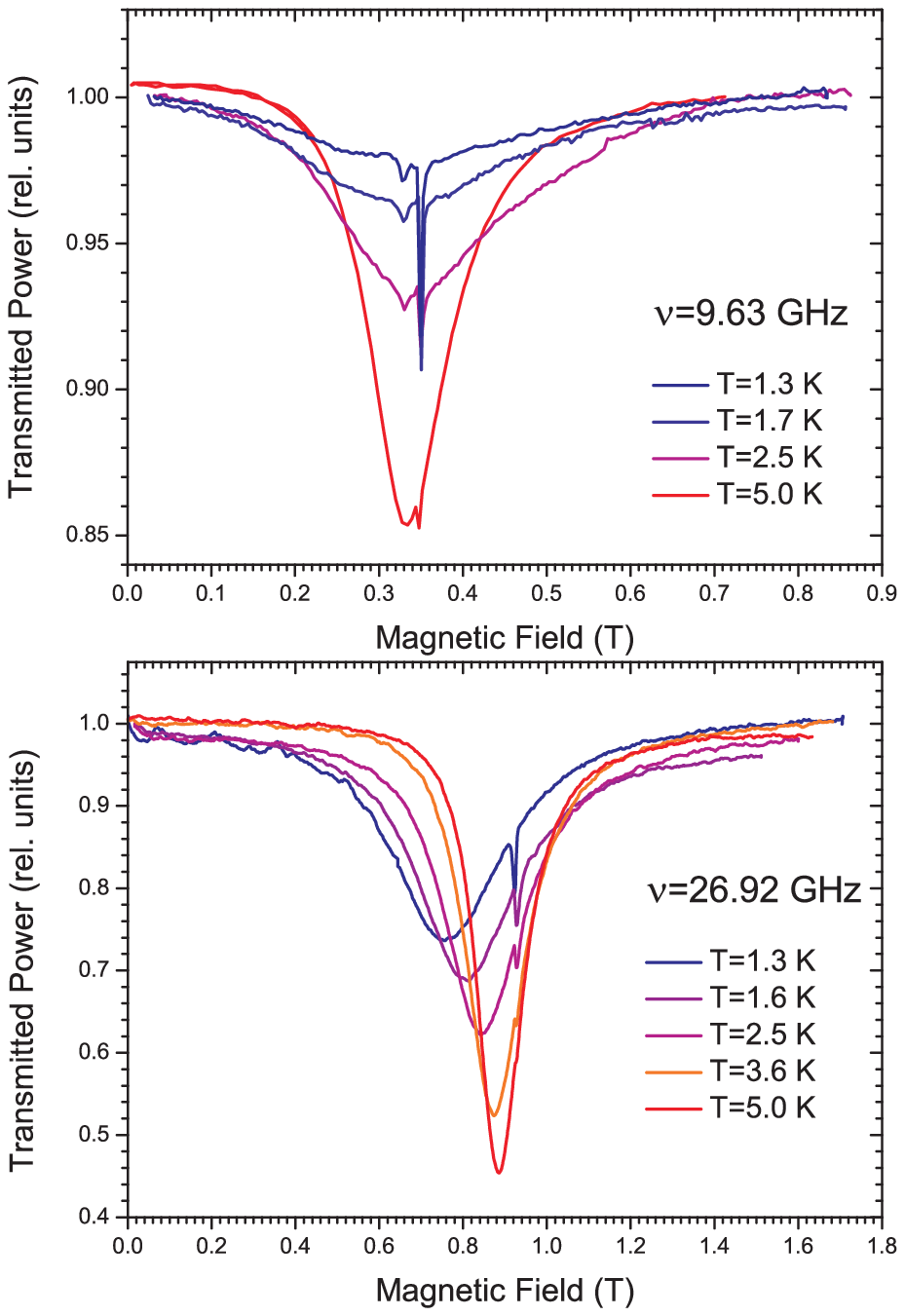}
    \caption{Upper panel: set of 9.63 GHz ESR lines at different temperatures for $H \parallel b$.
    The narrow resonance at $\sim 0.35$ T is the signal of g=2.0 DPPH marker;
    a tiny resonance, appearing only at low temperatures, is due to a paramagnetic impurities.
    Lower panel: set of 26.92 GHz ESR lines at different temperatures for $H \parallel b$.
    The narrow resonance at $\sim 0.9$ T is the DPPH signal.\label{b927Ghz}}
\end{figure}

We would like to describe here one possible reason for a pronounced frequency dependence of the ESR signal as
described in Section~\ref{sec:freq-T} above. Consider first ${\bf H}\parallel {\bf b}$ arrangement when our
analysis predicts that the signal is centered about the single frequency \be 2\pi \hbar ~\nu_{R/L} =
\sqrt{(g\mu_B H)^2 + \pi^2 (D_a^2 + D_c^2)/4}. \label{eq1} \ee This is just equation (1) of the main paper (see
also discussion following equations (4,5) there). This equation clearly states that there is a minimal ESR
frequency, given by $\nu_{\rm b,min}=\Delta/(2\pi \hbar) = D/(4\hbar)$, below which no resonance is possible in
this geometry. Experimentally the resonance is observed both above and below $\nu_{\rm b,min}$, but with
markedly different temperature dependencies. The low-frequency ($\nu < \nu_{\rm b,min}$) response remains
paramagnetic-like and quickly looses intensity on lowering $T$, see Figure~\ref{b927Ghz}. On the contrary, the
high-frequency response (Figure~\ref{b927Ghz}) follows prediction \eqref{eq1} rather nicely, as detailed in the
main text.

We'd like to argue that the observed difference is a finite-T effect. Eq.~\eqref{eq1} describes idealized $T=0$
situation of completely decoupled spin chains, to which \ccc\ is only an approximation in the intermediate
temperature window between $T_0$ discussed above and the ordering temperature $T_N$. In this temperature
interval one expects sizable thermal population of spinons at pretty much all momenta, not restricted to the
${\bf q}=0$ limit in which the current field theory is formulated. These thermal spinons are not particularly
correlated and can be thought of as forming an ideal gas of neutral fermions. It seems physically sensible to
think that their ESR response is paramagnetic and not sensitive to the low-energy effects due to the internal DM
field. It is this response that is measured at $\nu < \nu_{\rm b,min}$ in Fig.~\ref{b927Ghz}. According to this
argument the intensity of the paramagnetic part of the signal should diminish with lowering T, simply following
diminishing of the population of thermal spinons, in agreement with our observation in Fig.~\ref{b927Ghz}. The
higher-frequency response ($\nu > \nu_{\rm b,min}$) is instead the intrinsic property of the spin chain.

It is easy to extend this argument to other field directions. Let, for example, ${\bf H} \parallel {\bf a}$:
then according to equation (5) of the paper we can achieve the situation when $g_a\mu_B H_a = \pi D_a/2$ and
$2\pi\hbar \nu_R(z={\text{odd}}) = 2\pi\hbar \nu_L(z={\text{even}}) = \pi D_c/2$, while
$2\pi\hbar \nu_R(z={\text{even}}) = 2\pi\hbar \nu_L(z={\text{odd}}) = \pi\sqrt{D_c^2 + 4 D_a^2}/2$.
Thus in this case $\nu_{\rm a, min} = D_c/(4\hbar)$ and this condition is realized for $1/2$ of chains
in the system. Similar argument shows that for ${\bf H} \parallel {\bf c}$
the minimal resonance frequency is $\nu_{\rm c, min} = D_a/(4\hbar)$. To reiterate, below the corresponding
minimal resonance frequency we expect standard paramagnetic ESR signal due to thermally excited
population of spinons, and the intensity of such a signal should decrease with lowering of the temperature.

It is interesting to note that there is one `magic' orientation of the external field for which $\nu_{\rm min}=0$.
This happens for $g_a\mu_B H_a =  \pi D_a/2, H_b=0$ and $g_c \mu_B H_c = \pi D_c/2$, in which case
$1/4$ of the chains experiences zero total field. It would be interesting to study this
special orientation in more details in the future.